\documentclass[twocolumn]{aastex631}

\makeatletter
\newcommand\DefineObj[3][\@empty]{%
  \expandafter\newcommand\csname pkgwobj@l#2\endcsname{#3}%
  \ifx\@empty#1%
    \expandafter\newcommand\csname pkgwobj@s#2\endcsname{#3}%
  \else%
    \expandafter\newcommand\csname pkgwobj@s#2\endcsname{#1}%
  \fi}%
\newcommand{\pkgw@printobj@long}[1]{%
  \expandafter\ifx\csname pkgwobj@l#1\endcsname\relax%
    \textbf{[unknown object!]}%
  \else%
    \csname pkgwobj@l#1\endcsname%
  \fi}%
\newcommand{\pkgw@printobj@short}[1]{%
  \expandafter\ifx\csname pkgwobj@l#1\endcsname\relax%
    \textbf{[unknown object!]}%
  \else%
    \csname pkgwobj@s#1\endcsname%
  \fi}%
\newcommand{\obj}{\@ifstar{\pkgw@printobj@long}{\pkgw@printobj@short}}%
\makeatother

\DefineObj[2MASS~J0415$-$0935]{2m0415}{2MASS J04151954$-$0935066}

\begin{document}

\title{Diversity of Cold Worlds: A Near Complete Spectral Energy Distribution for 2MASS~J04151954$-$0935066 using JWST}
\shorttitle{SED for an Extrasolar World using JWST}
\shortauthors{Alejandro Merchan et al.}



\correspondingauthor{Sherelyn Alejandro Merchan}
\email{sherelyna12@gmail.com}

\author[0000-0003-0548-0093]{Sherelyn Alejandro Merchan}
\affiliation{Department of Astrophysics, American Museum of Natural History, New York, NY, USA}
\affiliation{Department of Physics, Graduate Center, City University of New York, New York, NY, USA}

\author[0000-0001-6251-0573]{Jacqueline K. Faherty}
\affiliation{Department of Astrophysics, American Museum of Natural History, New York, NY, USA}

\author[0000-0002-2011-4924]{Genaro Su\'arez}
\affiliation{Department of Astrophysics, American Museum of Natural History, New York, NY, USA}

\author[0000-0002-1821-0650]{Kelle L. Cruz}
\affiliation{Department of Physics and Astronomy, Hunter College, City University of New York, New York, NY, USA}
\affiliation{Department of Astrophysics, American Museum of Natural History, New York, NY, USA}
\affiliation{Department of Physics, Graduate Center, City University of New York, New York, NY, USA}

\author[0000-0002-6523-9536]{Adam J.\ Burgasser}
\affiliation{Department of Astronomy \& Astrophysics, UC San Diego, La Jolla, CA, USA}

\author[0000-0002-2592-9612]{Jonathan Gagn\'{e}}
\affiliation{Plan\'{e}tarium Rio Tinto Alcan, Montreal, Quebec, Canada}
\affiliation{D\'{e}partement de Physique, Universit\'{e} de Montr\'{e}al, Montreal, Quebec, Canada}

\author[0000-0003-1150-7889]{Callie E. Hood}
\affiliation{Department of Astronomy and Astrophysics, University of California, Santa Cruz, Santa Cruz, CA, USA}

\author[0000-0003-4636-6676]{Eileen C. Gonzales}
\affiliation{Department of Physics and Astronomy, San Francisco State University, San Francisco, CA, USA}

\author[0000-0001-8170-7072]{Daniella C. Bardalez Gagliuffi}
\affiliation{Department of Physics \& Astronomy, Amherst College, Amherst, MA, USA}
\affiliation{Department of Astrophysics, American Museum of Natural History, New York, NY, USA}

\author[0009-0009-3024-5846]{Jolie L'Heureux}
\affiliation{Department of Theoretical Physics and Astrophysics, Faculty of Science, Masaryk University, Brno, Czech Republic}

\author[0000-0003-0489-1528]{Johanna M. Vos}
\affiliation{School of Physics, Trinity College Dublin, The University of Dublin, Dublin, Ireland}

\author[0000-0002-6294-5937]{Adam C. Schneider}
\affiliation{United States Naval Observatory, Flagstaff, AZ, USA}

\author[0000-0002-1125-7384]{Aaron M. Meisner}
\affiliation{NSF National Optical-Infrared Astronomy Research Laboratory, Tucson, AZ, USA}

\author[0000-0002-4404-0456]{Caroline Morley}
\affiliation{Department of Astronomy, University of Texas at Austin, Austin, TX, USA}

\author[0000-0003-4269-260X]{J. Davy Kirkpatrick}
\affiliation{IPAC, Caltech, Pasadena, CA, USA}

\author[0000-0001-7519-1700]{Federico Marocco}
\affiliation{IPAC, Caltech, Pasadena, CA, USA}

\author[0000-0003-2102-3159]{Rocio Kiman}
\affiliation{Department of Astronomy, California Institute of Technology, Pasadena, CA, USA}

\author[0000-0002-5627-5471]{Charles A. Beichman}
\affiliation{IPAC, Caltech, Pasadena, CA, USA}

\author[0000-0003-4600-5627]{Ben Burningham}
\affiliation{Department of Physics, Astronomy and Mathematics, University of Hertfordshire, Hatfield, UK}

\author[0000-0001-7896-5791]{Dan Caselden}
\affiliation{Department of Astrophysics, American Museum of Natural History, New York, NY, USA}

\author{Peter R. Eisenhardt}
\affiliation{NASA Jet Propulsion Laboratory, California Institute of Technology, Pasadena, CA, USA}

\author[0000-0001-5072-4574]{Christopher R. Gelino}
\affiliation{IPAC, Caltech, Pasadena, CA, USA}

\author[0000-0002-4088-7262]{Ehsan Gharib-Nezhad}
\affiliation{NASA Ames Research Center, Mountain View, CA, USA}

\author[0000-0002-2387-5489]{Marc J. Kuchner}
\affiliation{Exoplanets and Stellar Astrophysics Laboratory, NASA Goddard Space Flight Center, Greenbelt, MD, USA}

\author[0000-0002-9420-4455]{Brianna Lacy}
\affiliation{Department of Astronomy, University of Texas at Austin, Austin, TX, USA}

\author[0000-0003-4083-9962]{Austin Rothermich}
\affiliation{Department of Astrophysics, American Museum of Natural History, New York, NY, USA}
\affiliation{Department of Physics, Graduate Center, City University of New York, New York, NY, USA}

\author[0000-0003-4225-6314]{Melanie J. Rowland}
\affiliation{Department of Astronomy, University of Texas at Austin, Austin, TX, USA}

\author[0000-0001-8818-1544]{Niall Whiteford}
\affiliation{Department of Astrophysics, American Museum of Natural History, New York, NY, USA}

\begin{abstract}

We present the a near complete spectral energy distribution (SED) for an extrasolar world: the T8 brown dwarf 2MASS~J04151954$-$0935066. Spanning from optical to mid-infrared (0.7--20.4~\micron) wavelengths, the SED for this substellar atmosphere is constructed from new JWST NIRSpec G395H ($R\sim$2700) and Magellan FIRE echelle ($R\sim$8000) near-infrared spectra, along with MIRI mid-infrared photometry complemented by spectra from Keck I, IRTF, Magellan, AKARI, Spitzer and photometry from various surveys and missions. The NIRSpec G395H spectrum reveals strong molecular absorptions from NH$_{3}$, CH$_{4}$, H$_{2}$S, CO$_{2}$ and H$_{2}$O at approximately 3.00, 3.35, 3.95, 4.25, and 5.00~\micron~respectively, along with the presence of a CO absorption feature detected mainly at $\sim 4.6~\micron$. We detect no absorption of near-infrared \ion{K}{1} doublets in the $R\sim8000$ FIRE spectra. In the mid-infrared IRS spectrum, we tentatively identify a new CO$_{2}$ feature at 14--16~\micron. The comprehensive SED allows us to empirically constrain bolometric luminosity, effective temperature, mass and radius. Additionally, we demonstrate that the NIRSpec G395H resolution, the highest allowable by JWST, enables a precise radial velocity measurement of $47.1\pm1.8$ km s$^{-1}$ for the object, in agreement with previous measurements. 

\end{abstract}



\section{Introduction}
The observations of Teide 1 \citep{rebolo_discovery_1995} and the cold companion \object{Gl~229B} \citep{Naka95,Oppe95} were the first detections of brown dwarfs.
With their discoveries, substellar mass physics began bridging the gap in our understanding between low mass stars and giant jovian worlds like those found in our solar system.
Brown dwarfs (objects with masses $<$ $\sim$ 75 M$_{Jup}$) were theorized in the 1960's (\citealt{Kumar63}) but required advancements in infrared surveys (i.e the Two-Micron All-Sky Survey \citealt{Cutri2003}) in order to be found in large quantities. Brown dwarfs range in temperature from $\sim$3000~K through $\sim$250K (e.g. \citealt{Kirkpatrick2024}) which correspond to the spectral classes of late-type M, L, T and Y dwarfs (\citealt{Martin1999}, \citealt{Kirk99}, \citealt{Burg2002}, \citealt{Cushing11}). Because the temperatures differ across spectral class, atmospheric chemistry hence molecular absorption in the observed data of brown dwarfs changes with spectral subtype.  For T-type dwarfs, water (H$_{2}$O), methane (CH$_{4}$), and ammonia (NH$_{3}$) gases can dominate the infrared.  Strong abundances of CO and CO$_{2}$ are also found in a variety of characterized T dwarfs (e.g \citealt{Yamamura2010, Mukh22,Miles2020}) which suggests an atmosphere in disequilibrium chemistry due to strong vertical mixing \citep{Grif99,Fegl96,Oppe95}. 

The majority of brown dwarf spectroscopic follow-up observations through the early 21st century were done with ground based facilities in the near infrared (e.g. \citealt{Burg2004}, \citealt{Kirkpatrick11,Kirkpatrick2010}) which covered $\sim$ $1 - 2.5 ~\micron$. However the 3 - 5 $~\micron$ spectral window contains critical coverage of CH$_{4}$, H$_{2}$O, NH$_{3}$ and bands of CO and CO$_{2}$ for brown dwarfs. Unfortunately, observations of this spectral region are challenging from the ground due to the brightness of the Earth's atmosphere. 
Fortunately, space missions like \textit{AKARI} collected spectroscopic observations probing this region of interest \citep{Muraka07}, furthering our understanding of exoplanets and brown dwarf atmospheres. The launch of the space-based infrared observatory the James Webb Space Telescope (JWST) enabled the continuation of these studies at a greater sensitivity \citep{Rigb23}. For instance, \citet{Miles23} presented a high fidelity 1--20~\micron~spectrum of the planetary companion late-type L dwarf VHS~J125601.92$-$125723.9 obtained with JWST. 
They illustrated how disequilibrium chemistry impacts the spectral shape, in addition to a vast array of molecular species. 

Discovered in \citet{Burg2002}, the T8 dwarf 2MASS~J04151954$-$0935066 (2MASS~J0415$-$0935 for short) has been the subject of several thorough studies and it is a standard for spectral classification of late-type, cool T dwarfs (e.g \citealt{Saum2007}, \citealt{Leggett2007}, \citealt{Miles2020}, \citealt{Yamamura2010}, \citealt{Hood2024}). Characterizing this object with additional, detailed JWST observations will solidify its use as a benchmark brown dwarf as it can be used to ground atmospheric models and comparative surveys of lesser known sources. 


In this paper we present new near- and mid-infrared observations for 2MASS~J0415-0935 complemented with archival ground and space based observations which yield a near complete (0.7--20.4~\micron) spectral energy distribution (SED) of an extrasolar atmosphere covering 93\% of its total bolometric luminosity. The data analyzed here-in precisely constrains the luminosity and effective temperature for 2MASS~J0415$-$0935. There are only a handful of similarly complete SEDs of brown dwarfs reported in the literature to date including (1) the aforementioned 0.97--19.8~\micron\ SED for VHS~1256~B in \citet{Miles23} which used JWST NIRSpec and MIRI spectroscopic observations, and (2) the 0.8--15~\micron\ SED for HN~Peg~B from \citet{Suar2021} which is composed of data from various space and ground based facilities.

In Section~\ref{DATA}, we present optical to mid-infrared spectrophotometry for 2MASS~J0415$-$0935. In Section~\ref{FUNDPARAMS}, we discuss the process of constructing the SED, how fundamental parameters were derived, and analyze the values in context with all known brown dwarfs. Additionally, we detail what is extracted from the data including kinematics and molecular features. Conclusions are presented in Section~\ref{CONCLU}.

\section{Observations and Data Reduction}\label{DATA}
2MASS~J0415$-$0935 has both new and archived data which are both discussed in detail in this section.  All data are available for download from the SIMPLE Archive\footnote{\url{http://simple-bd-archive.org/}}, a repository for data related to brown dwarfs, directly-imaged exoplanets, and low-mass stars.

\subsection{New Data}
We present new data for 2MASS~J0415$-$0935 acquired from the Magellan telescope and through the JWST Cycle~1 GO~\#2124 program (PI J. Faherty).

\subsubsection{Magellan/FIRE Echellete Spectrum}\label{FIRE}
\begin{figure*}[ht]
    \centering
    \includegraphics[width=\textwidth]{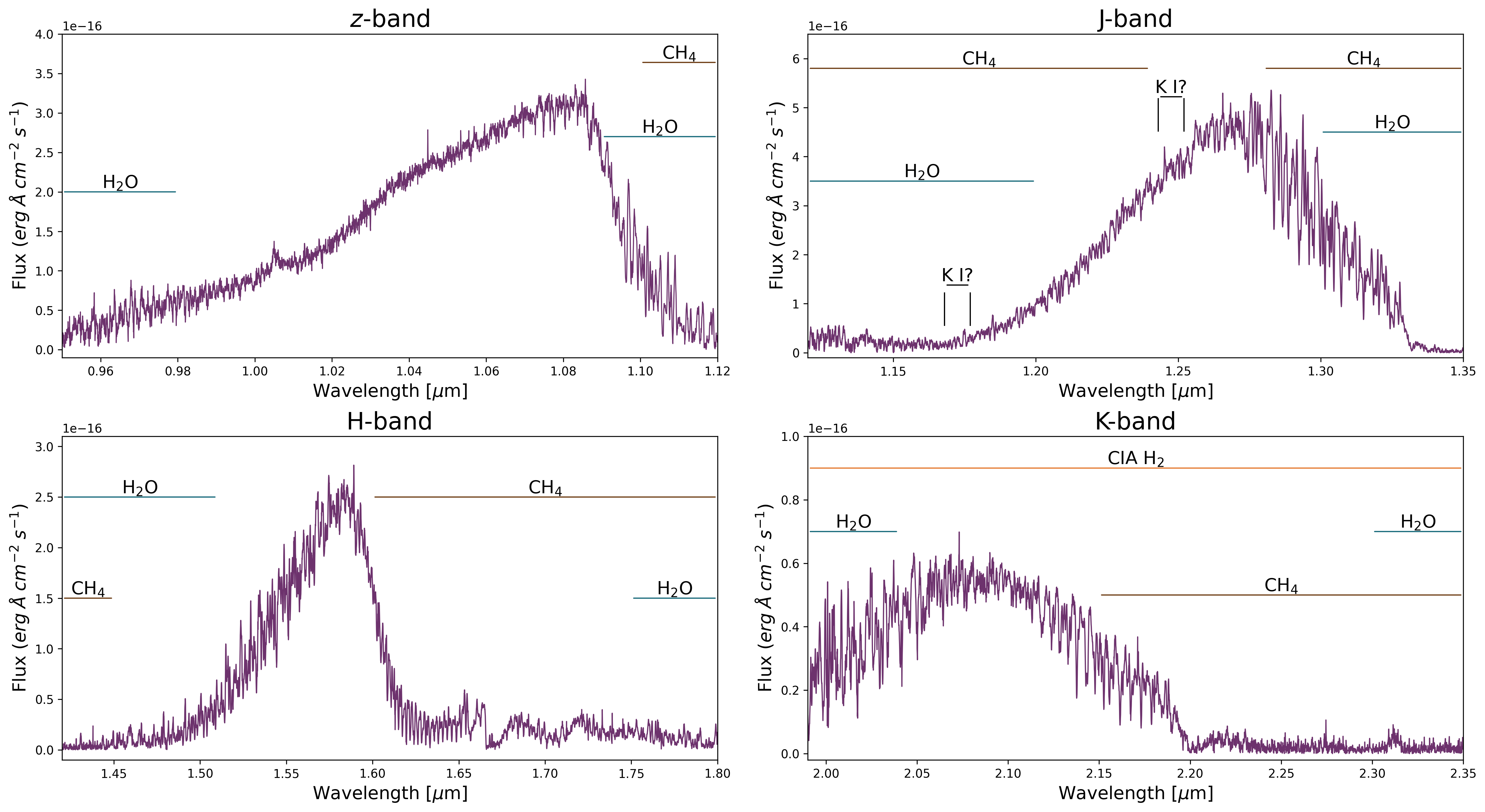}
    \caption{Magellan FIRE spectrum with panels showing the z-band, $J$-band, $H$-band, and $K$-band regions along with prominent molecular absorption from CH$_4$ and H$_2$O.}
    \label{fig:fire_zjhk}
\end{figure*}

During 2010 September 20 (UT), we acquired near-infrared spectroscopic observations of 2MASS~J0415$-$0935 using the Folded-port InfraRed Echellete (FIRE) spectrograph on the 6.5m Baade Magellan telescope  \citep{Simc2013}. Observing conditions were clear with $0.5\arcsec$ seeing. We used echellete mode and a $0.6\arcsec$ slit with $2\times900\sec$ exposures (AB) at an average airmass of 1.063. The A0~V star HD~31004 was observed for flux calibration. The data were reduced using the FIREHOSE pipeline which is based on the MASE and Spextool reduction packages \citep{Vacca2003,Cushing2004,Bochanski2009}\footnote {\url{https://wikis.mit.edu/confluence/display/FIRE/FIRE+Data+Reduction}}. 
The resulting spectrum has a wavelength coverage of $0.8-2.5~\micron$ with a resolution $R\sim$ 8000 and is shown in Figure~\ref{fig:fire_zjhk}.

\subsubsection{JWST NIRSpec G395H medium-resolution spectroscopy}
NIRSpec data for 2MASS~J0415$-$0935 were obtained on 16~October~2022 using the F290LP filter, the G395H grating, the S200A1 aperture and the SUB2048 subarray. Acquisition images were first obtained for each target using the WATA method, the CLEAR filter, and the NRSRAPID readout pattern. We used 11 groups per integration, 3 integrations per exposure and 3 total dithers for a summation of 9 total integrations in 168.488 seconds of exposure time.  

\begin{figure*}
    \includegraphics[width=\textwidth]{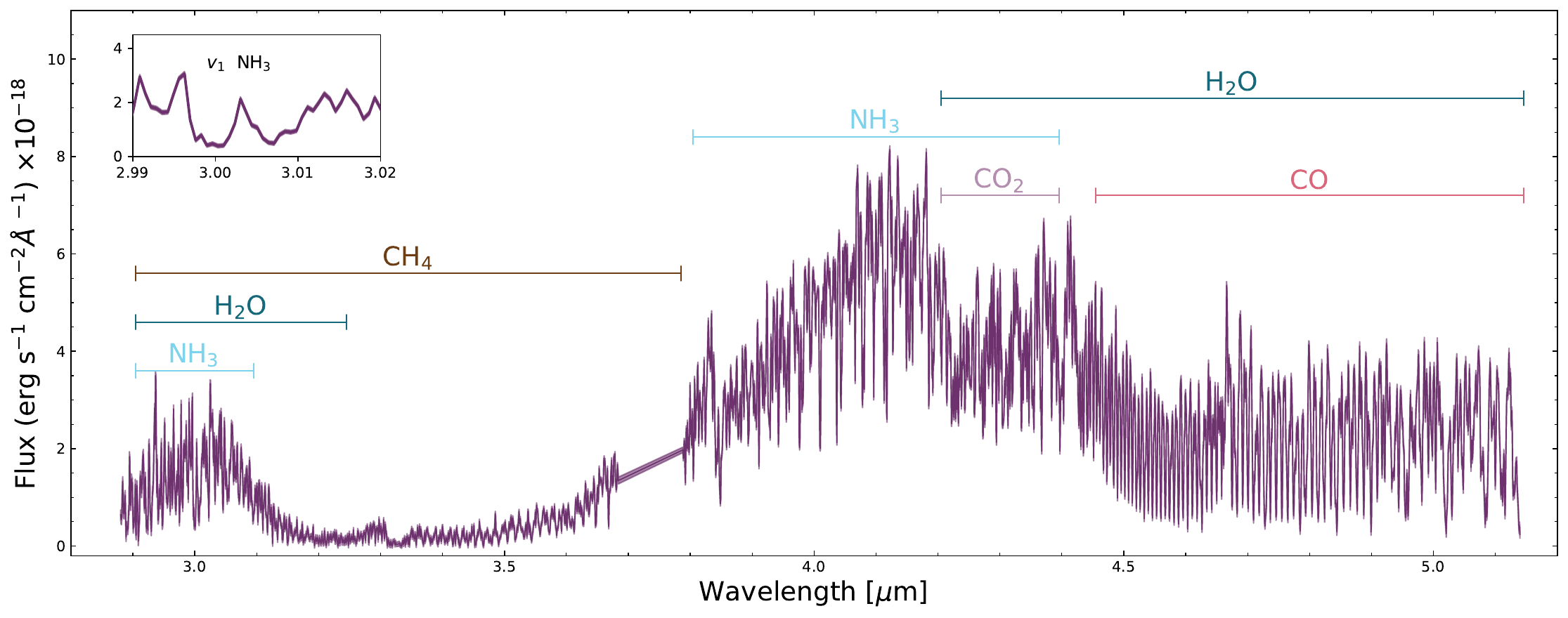}
    \caption{JWST NIRSpec G395H spectrum for 2MASS J0415$-$0935 with the uncertainties in the shaded region. The inset plot displays a zoom into the $v_{1}$  NH$_3$ band.}
    \label{fig:xsec_nh}
\end{figure*}

For the reduction of the NIRSpec G395H spectra, we ran the JWST calibration pipeline v1.10.0, using the Calibration Reference Data System (CRDS) context file jwst\_1146.pmap, and default parameters. 
We optimized the aperture extraction considering the slit position of the target.
The resulting combined spectrum has a wavelength coverage that ranges from 2.87 to 5.14~$\micron$ with a resolution of $\sim$2700 and is shown in Figure~\ref{fig:xsec_nh}.

\subsubsection{JWST MIRI photometry}
MIRI photometry was obtained on 18 September 2022 with the F1000W (9.023--10.891~\micron), F1280W (11.588--14.115~\micron), and F1800W (16.519--19.502~\micron) filters.  For each filter the FASTR1 readout pattern was chosen with a 2-point dither pattern.  2MASS J0415$-$0935 was observed with MIRI using 5 groups per integration for each filter.  Total exposure time plus overhead for the MIRI observing of 2MASS J0415$-$0935 was 0.52 hours. 

We used the MIRI photometry provided by the JWST pipeline, which agrees well with the existing Spitzer IRS spectrum.
We calculated synthetic F1000W, F1280W, and F1800W fluxes from the Spitzer IRS spectrum. 
The flux differences between the MIRI photometry and IRS synthetic fluxes are well within the uncertainties
($1.703\pm0.022$~mJy, $1.393\pm0.038$~mJy, and $0.708\pm0.264$~mJy) in the F1000W, F1280W, and F1800W filters, respectively. 
The IRS Spectrum does not fully cover the F1800W filter, so the reported flux difference is an upper value. 

\subsection{Literature Data}
In order to construct the SED of 2MASS~J0415$-$0935, we complemented the new data with previously published data which is publicly available in the SIMPLE Archive.

\subsubsection{Keck I/LRIS Spectrum}
\citet{Burg2003} observed 2MASS~J0415$-$0935 with the Low Resolution Imaging Spectrometer (LRIS) mounted on the Keck~I 10m telescope \citep{Oke1995}, spanning the 0.63--1.01~\micron~ wavelength range with a resolution of R$\sim$1000. 


\subsubsection{IRTF/SpeX Spectrum}
\citet{Burg2004} published a near-infrared ($0.8-2.5\,{\micron}$), low resolution (R $\sim$ 150) spectrum of 2MASS~J0415$-$0935, observed using the SpeX spectrograph mounted on the Infrared Telescope Facility (IRTF) on Mauna Kea, HI \citep{Rayner2003}. 2MASS~J0415$-$0935 was designated as a late-T spectral standard and consequently aided in the classification of subsequent T dwarf discoveries.

\subsubsection{AKARI/IRC Spectrum}
\citet{Yamamura2010} published mid-infrared spectroscopic observations of L1--T8 dwarfs with the InfraRed Camera from the $AKARI$ mission \citep{Muraka07} with wavelength coverage of $2.5-5~\micron$ and a spectral resolution of R$\sim$120. 2MASS~J0415$-$0935 was one of the targets published with AKARI data. \citet{Yamamura2010} examined the strength of molecular absorption bands from CH$_{4}$ at $3.3~\micron$, CO$_{2}$ at $4.2~\micron$, and CO at $4.6~\micron$.

\subsubsection{Spitzer/IRS Spectrum}
\citet{Suar2022} reprocessed a mid-infrared (5.2--20~$\micron$) low-resolution ($R\sim100$) spectrum of 2MASS~J0415$-$0935 obtained with the Infrared Spectrograph (IRS) on the Spitzer Space Telescope \citep{Houck2004}, originally published in \citet{Saum2007}.

\begin{figure*}
   \centering
   \includegraphics[scale=.41]{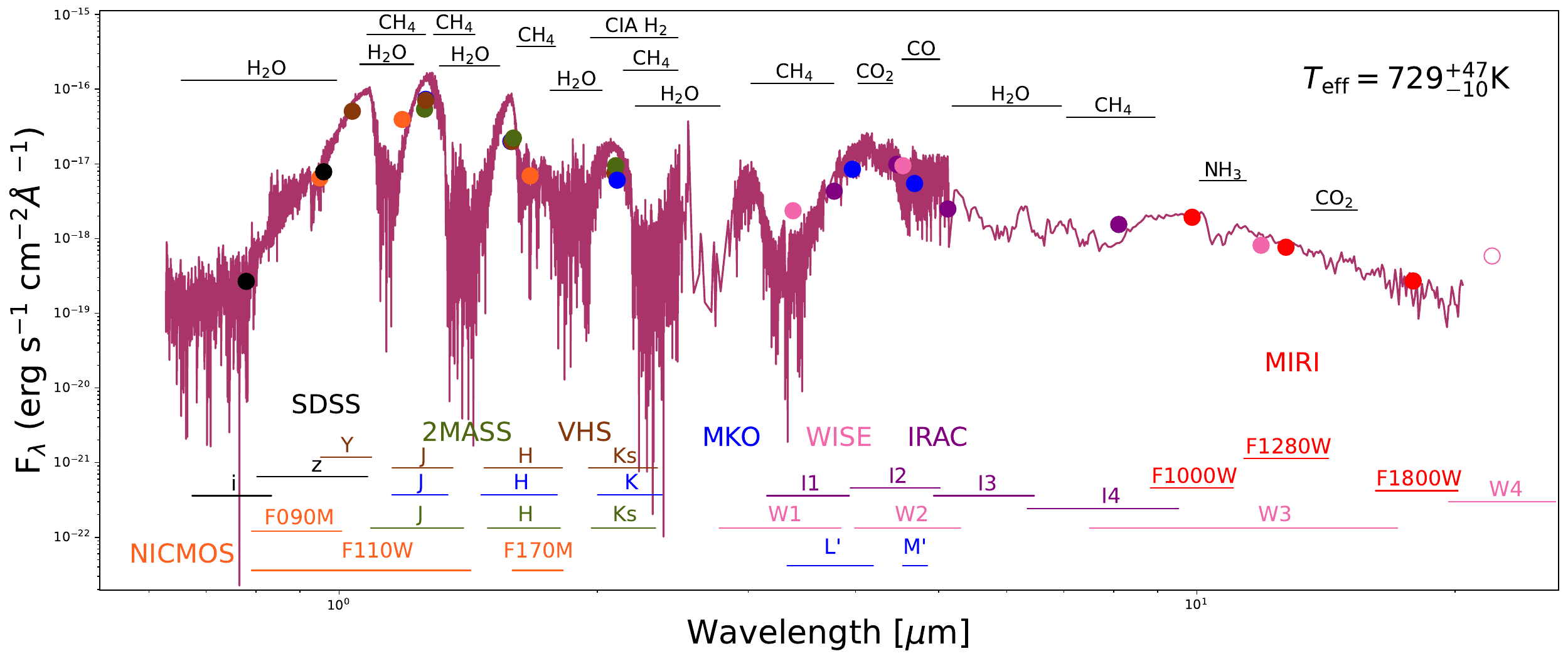}
   \caption{Distance calibrated SED of 2MASS~J0415$-$0935 using new JWST NIRSpec and MIRI observations and spectrophotometry from the literature, as indicated in the plot. We label the principal molecular bands for H$_{2}$O, CH$_{4}$, CIA H$_{2}$, NH$_{3}$ and CO along with the calculated $T_{\mathrm{eff}}$ of the object. The photometry used is denoted by circles in the same corresponding color to the filters shown at the bottom of the plot and the circles are centered at the effective wavelength of each filter, which were obtained using the SED for 2MASS~J0415$-$0935.}
   \label{fig:0415_full_sed}
\end{figure*}

\subsubsection{Literature Photometric Observations}
There are a number of publicly available photometric observations for 2MASS~J0415$-$0935 that also contribute to the SED. For the optical photometry, magnitudes published in \citet{Legg2012} for the $i$ and $z$ filters from the Sloan Digital Sky Survey \citep[SDSS; ][]{York00} were used. Offsets from \citet{Hewett2006} were implemented to convert values from the AB to Vega system. For near-infrared photometry, we used data from the Two Micron All Sky Survey (2MASS), the Vista Hemisphere Survey (VHS) Data Release 6, and the UKIRT Fast-Track Imager (UFTI) on the United Kingdom Infrared Telescope \citep{Knapp2004,Lodieu2012,Golim2004}. The latter uses the Mauna Kea Observatory filters. Additionally, photometry in the three filters F090M, F110W and F170M from the Near Infrared Camera and Multi-Object Spectrometer (NICMOS) on the \textit{Hubble Space Telescope} published in \citet{Burg2006b} was included. Photometry at 3.6~\micron (ch1), 4.5~\micron (ch2), 5.7~\micron (ch3), and 7.8~\micron (ch4) from the InfraRed Array Camera (IRAC) on the \textit{Spitzer Space Telescope} was also included \citep{Patt2006}. Mid-infrared magnitudes from the Wide Field Infrared Survey Explorer (WISE) were also included. WISE W1 and W2 photometry were taken from the CatWISE2020 catalog \citet{Maroc21} while W3 and W4 bands were from AllWISE \citep{Cutri2013}. All these observations are listed in Table~\ref{tab:phot}. 

\begin{deluxetable}{lcc}
    \tablecaption{Photometry of 2MASS J0415$-$0935\label{tab:phot}}
    \startdata
    \tablehead{
    \colhead{Band} & \colhead{Vega magnitude} & \colhead{Reference}}
    SDSS $i$      & $23.084\pm0.090$ & L12  \\
    SDSS $z$      & $18.867\pm0.090$ & L12  \\
    2MASS $J$     & $15.695\pm0.058$ & C03  \\
    2MASS $H$     & $15.537\pm0.113$ & C03  \\
    2MASS $K_S$   & $15.429\pm0.201$ & C03  \\
    VHS $Y$       & $16.451\pm0.008$ & VDR6 \\
    VHS $J$       & $15.327\pm0.004$ & VDR6 \\
    VHS $H$       & $15.680\pm0.012$ & VDR6 \\
    VHS $K_S$     & $15.683\pm0.023$ & VDR6 \\
    MKO $J$       & $15.32\pm0.03$   & K04  \\
    MKO $H$       & $15.70\pm0.03$   & K04 \\
    MKO $K$       & $15.83\pm0.03$   & K04 \\
    MKO $L'$      & $13.28\pm0.05$   & G04 \\
    MKO $M'$      & $12.82\pm0.15$   & G04 \\
    HST $F090M$   & $19.04\pm0.10$   & B06  \\
    HST $F110W$   & $16.47\pm0.05$   & B06  \\
    HST $F170M$   & $16.67\pm0.06$   & B06  \\
    WISE $W1$     & $15.148\pm0.026$ & M21 \\
    WISE $W2$     & $12.305\pm0.011$ & M21 \\
    WISE $W3$     & $11.132\pm0.113$ & C13 \\
    WISE $W4$     & $>8.638$         & C13 \\ 
    Spitzer [3.6] & $14.256\pm0.019$ & K19 \\    
    Spitzer [4.5] & $12.374\pm0.017$ & K19 \\
    Spitzer [5.7] & $12.87\pm0.070$  & P06 \\
    Spitzer [7.8] & $12.11\pm0.050$  & P06 \\
    JWST $F1000W$ & $10.749\pm0.002$ & TW  \\
    JWST $F1280W$ & $10.669\pm0.002$ & TW  \\
    JWST $F1800W$ & $10.327\pm0.007$ & TW  \\
    \enddata
    \tablerefs{C03: \citet{Cutri2003}; D12: \citet{Dupu2012}; H21: \citet{Hsu2021}; B03: \citet{Burg2003}; L12: \citet{Legg2012}; VDR6: VHS Data Release 6 ; B21: \citet{Best2021}; K04: \citet{Knapp2004}; G04: \citet{Golim2004}; B06: \citet{Burg2006b}; M21: \citet{Maroc21}; C13: \citet{Cutri2013}; K19: \citet{Kirk2019}; P06: \citet{Patt2006}; TW: This Work}
\end{deluxetable}

 \begin{figure*}
   \centering
   \includegraphics[width=\linewidth]{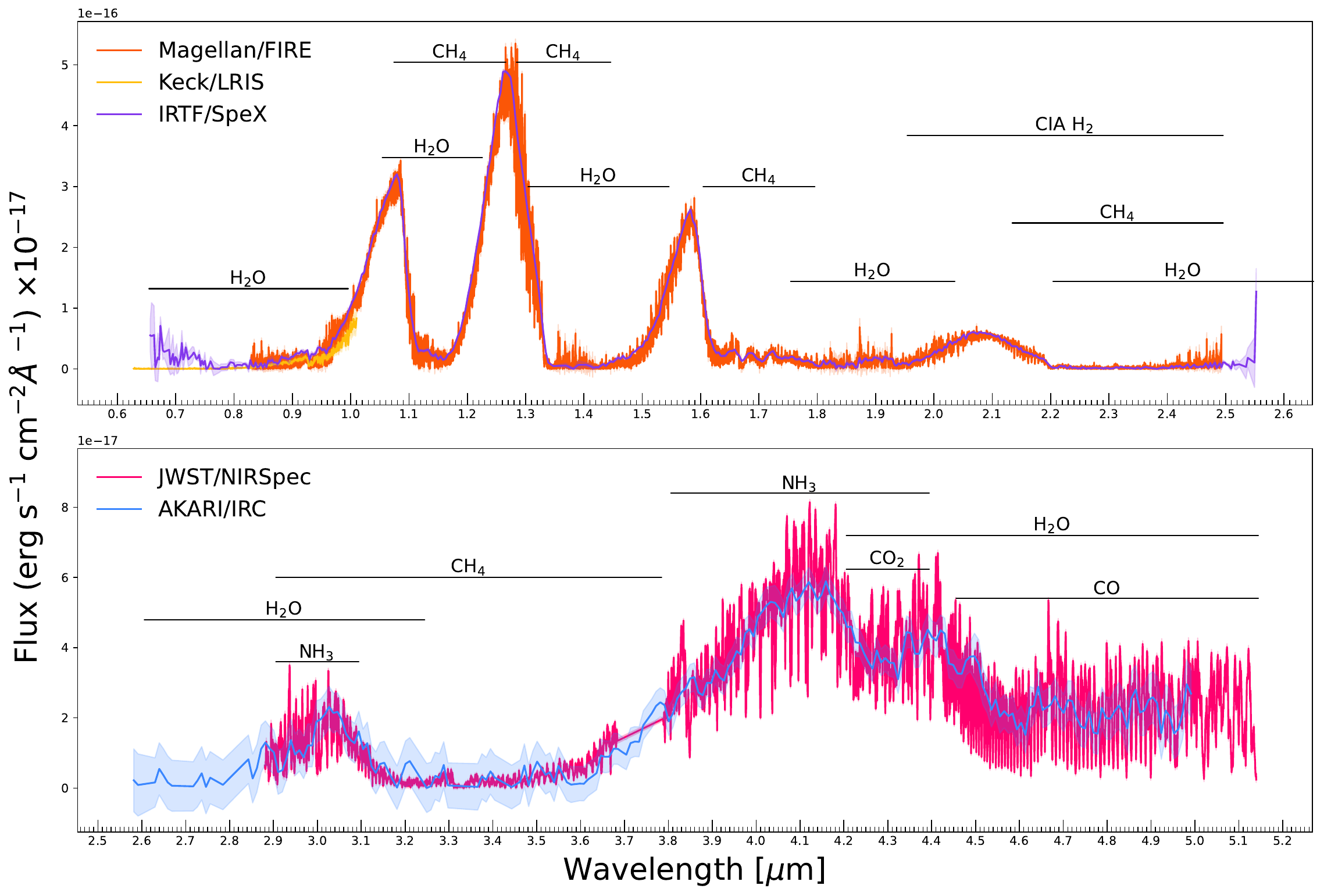}
   \caption{\textbf{Top panel:} Spectra from Magellan/FIRE (orange), Keck/LRIS (yellow), and IRTF/SpeX (purple). \textbf{Bottom panel:} Spectra from JWST/NIRSpec (pink) and AKARI/IRC (blue). Shaded regions represent the flux uncertainties of each individual spectrum. The main spectral features are indicated.}
   \label{fig:0415_overlap}
\end{figure*}

\section{Spectral Energy Distribution \& Fundamental Parameters of 2MASS J0415-0935}\label{FUNDPARAMS}

\subsection{Constructing the SED}\label{CSED}

The open-source Python package SEDkit V.2.0.5\footnote{V.2.0.5 of SEDkit fixed bugs in previous versions related to uncertainties in the fundamental parameters and calculations of the absolute magnitudes.} \citep{filippazzo_v2.0.5.sedkit_2024} was used in order to facilitate the construction of the SED shown in Figure~\ref{fig:0415_full_sed}. SEDkit was used to calculate a precise bolometric luminosity ($L_{\rm bol}$) and semi-empirically determine effective temperature along with other fundamental parameters. It was modified to include a Monte-Carlo approach (Section~\ref{sec:monte_carlo_approach}) during calculations to work in conjunction with an open-source python wrapper called SEDkitSIMPLE\footnote{\url{https://github.com/dr-rodriguez/SEDkitSIMPLE}}.   The wrapper was developed to automate the process of loading data from the SIMPLE Archive into SEDkit. 

The process of SED construction was as follows: 
\begin{description}
    \item[Step 1] We used SEDkit to call SIMPLE and load the spectra and photometry presented in Section~\ref{DATA}. 
    \item[Step 2] The published effective wavelengths for each filter were determined assuming a Vega spectrum \citep{SVO2020}. In order to better represent these photometry values in the context of the entire SED, we calculated the effective wavelengths using the actual spectra of 2MASS~J0415$-$0935. These effective wavelengths are used to plot the photometry points shown in Figure~\ref{fig:0415_full_sed} and are included as Data Behind the Figure (DbF).
    \item[Step 3] Spectra with overlapping regions are shown in Figure~\ref{fig:0415_overlap}. With the exception of the AKARI and NIRSpec spectra, overlapping spectra were combined by smoothing the higher signal to noise (SNR) spectrum to the lowest resolution wavelength array and then normalized following \cite{Fili2015}. In the 2.9--5~\micron\ region, the AKARI spectrum was trimmed to give preference to the JWST observations. 
    \item[Step 4] The spectra and photometry were distance-scaled using a parallax of $175.2\pm1.7$~mas \citep{Dupu2012}.
    \item[Step 5] SEDkit constructed the distance-scaled flux-calibrated SED of our object by starting at zero in wavelength and then linearly interpolating to the shortest wavelength data point (0.7~\micron), continuing with the observations, 
and then appending a Rayleigh-Jeans tail from the longest available wavelength data point at 20.4~\micron\ up to 1000~\micron. The flux from the SED shown in Fig~\ref{fig:0415_full_sed} was integrated to provide a measurement of the $L_{\rm bol}$. 
    \item[Step 6] In order to calculate the effective temperature ($T_{\mathrm{eff}}$) using the Stefan-Boltzmann law and the calculated $L_{\rm bol}$, a radius is required.  We estimated a radius using the best age approximation and the solar metallicity, hybrid cloud evolutionary model isochrones from \citet{Saumon2008}. 2MASS~J0415$-$0935 shows no signs of low gravity, nor low metallicity but rather has kinematics consistent with a field age brown dwarf (see for e.g \citealt{Hsu2021}, \citealt{Hood2024}).  Therefore we assumed a broad age range of 0.5--8.5~Gyr to semi-empirically extract all other fundamental parameters.
    \item[Step 7] SEDkit returns an SED for 2MASS~J0415$-$0935 and all fundamental parameters. 
\end{description}
All parameters calculated are presented in Table~\ref{tab:params} and the distance-calibrated SED is shown in Figure~\ref{fig:0415_full_sed}. 

\begin{deluxetable}{lcc}
    \tablecaption{Fundamental Parameters of 2MASS J0415$-$0935\label{tab:params}}
    \startdata
    \tablehead{
    \colhead{Parameter} & \colhead{value} & \colhead{unit} }
    log($L_{\rm bol}/{L_\odot}$) & $-5.71_{-0.01}^{+0.01}$ &     \\
    $T_{\mathrm{eff}}$           & $729_{-10}^{+47}$       &  K  \\
    Radius  & $0.855_{-0.020}^{+0.110}$ & $R_{\mathrm{Jup}}$  \\
    Mass    & $37_{-12}^{+10}$           & $M_{\mathrm{Jup}}$  \\
    $\log g$  & $5.13_{-0.35}^{+0.12}$  & dex  \\ 	
    \enddata
\end{deluxetable}

\subsection{Monte-Carlo Approach for Uncertainty Estimates} 
\label{sec:monte_carlo_approach}
The V.2.0.5  of SEDkit used a Monte-Carlo approach to derive model-dependent fundamental parameters and uncertainties. This Monte-Carlo approach used the given age range of the object along with the calculated $L_{\rm bol}$ and its uncertainty. It then generated a Gaussian distribution for bolometric luminosity using the $L_{\rm{bol}}$ value as the peak and its uncertainty as the standard deviation. For the age distribution of the object, we assumed a uniform distribution between the lowest (0.5~Gyr) and highest (8.5~Gyr) end of our adopted age range . We sampled $10^4$ values from these distributions and combined each $L_{\rm{bol}}$-age pair into the evolutionary models to acquire $M$, $R$, and $\log g$ values. We then considered the $68\%$ confidence interval  as the uncertainties on the model-dependent parameters, and took the median as the parameter values provided.

\subsection{Analysis of Fundamental Parameters}
Comparing the fundamental parameter values listed in Table~\ref{tab:params} to those in previous results, namely \citet{Hood2024,Zales22,Fili2015,Saum2007,Burg2006a,Vrba2004,Golim2004}, we find they generally agree within $2\sigma$. For instance, the work presented in \citet{Fili2015} used the same method of determining fundamental parameters described in section 3.1 -- excluding the aforementioned changes to the code -- making it a good comparison point. Nonetheless, there are several differences that are of important to note including a change in parallax measurement, and an addition of mid-infrared spectra (AKARI, JWST, and Spitzer) and photometry (VHS, HST, and MIRI). The $L_{\rm bol}$ from this paper is slightly larger than the calculated value in \citet{Fili2015} although it is within $3\sigma$. Similarly, our $T_{\rm eff}$ is within $1\sigma$ from \citet{Fili2015}, yet it is slightly warmer given that our SED has mid-infrared observations missing from the \citet{Fili2015} analysis.  


The work from \citet{Hood2024} conducts a spectral inversion analysis of this source and makes use of the SED presented in this paper exclusive of the Magellan, LRIS, and AKARI spectra, making it the most comparable. As noted in \citet{Hood2024}, the retrieved fundamental parameters are within $1\sigma$ of the SED analysis, with the exception of their smaller radius which is $3\sigma$ lower than our results. This difference is noted by \citet{Hood2024}, as being a standard problem in brown dwarf retrieval studies. 

Other works like those presented in \citet{Vrba2004} introduced a $T_{\mathrm{eff}}$ of $764^{+88}_{-71}$~K for 2MASS~J0415$-$0935, which is within $1\sigma$ of our work. By using their $M_{\rm bol}$ estimate of $18.70\pm{0.26}$ mag they derived an $L_{\rm bol}$ of $-5.58\pm{0.10}$~$L_\odot$, which is within $2\sigma$ from our results. Similarly, \citet{Golim2004} presented a range of 600--700~K for the $T_{\mathrm{eff}}$ of 2MASS~J0415$-$0935 based on a parallax measurement plus near-infrared spectra and photometry. The limited spectral coverage resulted in a temperature $3\sigma$ cooler than our results. These comparisons demonstrate the need for full spectral coverage to account for substellar object flux and precisely measure fundamental parameters.

The observed and estimated properties of 2MASS~J0415$-$0935 are provided in Table~\ref{tab:params}.

\subsection{Radial Velocity}\label{RV}
With the high resolution of the G395H spectra, we were able to calculate a radial velocity for 2MASS~J0415$-$0935.  We used a forward model sampling method with MCMC using four parameters (radial velocity, lsf width, 2 blaze parameters) on many segments of the spectrum, adopting the segments  with the most stable and reproducible RVs in each model grid. In our approach we compared to the \citet{Lacy23} models, the Elf Owl models from \citet{Mukherjee24}, and the \citet{Morley2012} models. We obtained a radial velocity measurement for 2MASS~J0415$-$0935 of $47.1\pm1.8$ km s$^{-1}$. Our results are within $2\sigma$ from the RV of $51.1\pm1.8$ km s$^{-1}$ published in \citet{Hsu2021} using Keck NIRSpec data.

\subsection{Kinematic Analysis}
Using the proper motion, parallax, and radial velocity we calculated component velocities for 2MASS~J0415$-$0935. We found U, V, and W values of $-63\pm1.0$, $-42.4\pm 0.6$, and $17\pm1.0$~km s$^{-1}$ respectively with a total velocity of $77.8\pm1.5$ km s$^{-1}$. Using the BANYAN~$\Sigma$ tool \citep{Gagne18}, we input all astrometric parameters and evaluated if 2MASS~J0415$-$0935 shows any kinematic correlation with a known moving group or association. BANYAN~$\Sigma$ revealed a 99.9\% field population probability, re-affirming that it is a field dwarf.

\begin{deluxetable}{lclc}
    \tablecaption{Astrometric and Kinematic properties of 2MASS J0415$-$0935\label{tab:uvw}}
    \startdata
    \tablehead{
    \colhead{parameter} & \colhead{value} &  \colhead{Units} &\colhead{Reference}}
    R.A. (J2000)    & 04:15:19.54008   &      &  C03    \\
    Decl. (J2000)   & -09:35:06.6012   &      &  C03     \\
    $\mu_{\alpha}$ cos $\delta$  &  $2214.3\pm1.2$&   mas yr$^{-1}$ & D12  \\
    $\mu_{\delta}$ & $535.9\pm1.2$  &   mas yr$^{-1}$ & D12 \\
    Parallax    &  $175.2\pm1.7$ &   mas           & D12  \\
    $v\sin i$   &   $33.5\pm2.0$ &   km s$^{-1}$   & H21  \\
    X & $-4.09\pm0.04$ &     pc & TW\\
    Y & $-1.73\pm0.02$ &     pc & TW \\
    Z & $-3.59\pm0.03$ &     pc & TW\\
    U  & $-63\pm1.00$   &   km s$^{-1}$ & TW \\
    V  & $-42.4\pm0.60$ &   km s$^{-1}$ & TW \\
    W  & $17.0\pm1.00$  &   km s$^{-1}$ & TW \\
    radial velocity & $47.1\pm1.80$  &   km s$^{-1}$ & TW\\
    Total Velocity & $77.8\pm1.50$ &  km s$^{-1}$ & TW \\
    \enddata
    \tablerefs{C03: \citet{Cutri2003}; D12: \citet{Dupu2012}; H21: \citet{Hsu2021}; B03: \citet{Burg2003}; L12: \citet{Legg2012}; VDR6: VHS Data Release 6 ; B21: \citet{Best2021}; K04: \citet{Knapp2004}; G04: \citet{Golim2004}; B06: \citet{Burg2006b}; M21: \citet{Maroc21}; C13: \citet{Cutri2013}; K19: \citet{Kirk2019}; P06: \citet{Patt2006}; TW: This Work}
\end{deluxetable}

\subsection{Spectral Features}
Within this section we describe spectral features we observed in 2MASS~J0415$-$0935.

\subsubsection{Optical}
The optical portion of the SED was previously published in the work of \citet{Burg2003}. The LRIS spectrum 
displays molecular absorption features such as FeH, CrH, \ion{K}{1}, and H$_{2}$O. \citet{Burg2003} also found an absence of any \ion{Li}{1} absorption in 2MASS~J0415$-$0935.  This might be due to the lower signal to noise of the optical data or a very small absorption abundance of \ion{Li}{1} in this object.  To date only a few T-dwarfs have yielded a \ion{Li}{1} detection \citep{Faherty2014,Pineda2016,Martin2022}.
\begin{figure*}[t]
   \centering
   \includegraphics[scale=0.65]{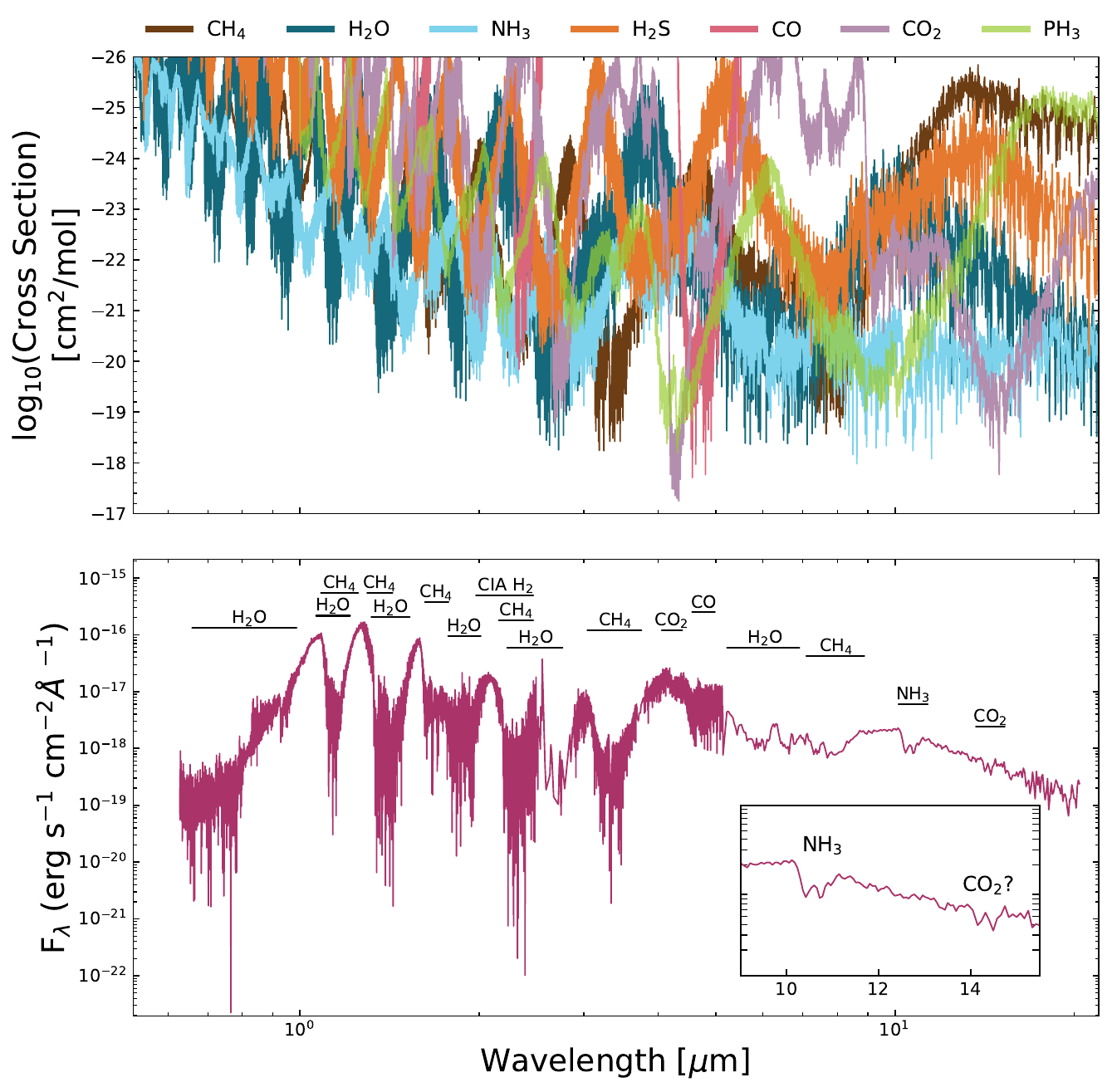}
   \caption{{\bf Top Panel:} Absorption cross sections for a temperature of 650K and pressure of 1 bar used in the retrieval analysis from 
   \citet{Hood2024}. {\bf Bottom Panel:} The complete SED for 2MASS~J0415$-$0935 with an inset showing the 9 - 15 $\micron$ region. Prominent absorption features are labeled including a tentative CO$_{2}$ feature between 14 - 16 $\micron$.}
   \label{fig:co_feature}
\end{figure*}

\subsubsection{Near-Infrared and Mid-Infrared}
Typically, near-infrared spectroscopic observations of T dwarfs are demarcated by the presence of H$_{2}$O and CH$_{4}$ in the $J$, $H$, and $K$~bands, along with collision induced absorption (CIA) due to H$_{2}$ in the $K$ band \citep{Miles2020,Burg2002,Legg2000}.  
In Figure~\ref{fig:fire_zjhk}, the $z$, $J$, $H$, and $K$~bands from the FIRE spectrum are displayed. Absorption features from CH$_{4}$ and H$_{2}$O are labeled in the z-band (0.95--1.12~\micron) panel. In the $J$-band (1.12--1.35~\micron) panel, absorption features from H$_{2}$O and CH$_{4}$ are clearly identified. The location of the K~I doublet at 1.243/1.252 and 1.168/1.177 $\micron$ is labeled, but not detected in the $R\sim8000$ FIRE spectrum for this late-T dwarf. Given the median SNR of $\sim 98$ in this region and the overall resolution of $R\sim$ 8000, the \ion{K}{1} doublet should be detectable if present, but it is absent. Like the previous two bands, in the $H$~band (1.45--1.80~\micron), CH$_{4}$ absorption is prominent. Lastly, the $K$-band (2.0--2.35~\micron), is marked by the presence of CIA H$_{2}$, H$_{2}$O, and CH$_{4}$. A CO absorption band centered at 2.3~\micron~ is commonly observed in warmer M, L, and early T dwarfs \citep{Burg2002,Legg2000}, however is not present in 2MASS~J0415$-$0935. As has been found in late-type T type dwarfs, methane absorption overpowers this region of the spectrum. 

The JWST NIRSpec spectrum shown in Figure~\ref{fig:xsec_nh} displays notable features from H$_{2}$O, CH$_{4}$, CO$_{2}$ and CO. A strong CO feature at 4.6--5.0~\micron~clearly demonstrates the CO disequilibrium chemistry --which was observed in brown dwarfs with the discovery of Gl~229B \citep{Naka95,Oppe95} -- and confirms the result seen in the lower resolution, lower SNR AKARI spectrum \citep{Yamamura2010}.

Within the NIRSpec spectrum, we find the $\upsilon_{1}$ band feature of NH$_{3}$ at 3$~\micron$. This feature was tentatively identified for the first time in a cold substellar object in the NIRSpec low-resolution data of the Y0 dwarf WISE~J035934.06-540154.6 (\citealt{Beil2023}). In the atmospheric retrieval work of \citet{Hood2024}, a total abundance value of log(NH$_{3}$)=-5.0$^{+0.04}_{-0.03}$ was retrieved using the SpeX prism, JWST NIRSpec, and Spitzer IRS spectra.  Our visual inspection complemented by the \citet{Hood2024} value confirms the presence of the NH$_{3}$ feature. 

 To be thorough in our examination, we checked for the presence of lithium chloride (LiCl) in the near and mid-infrared portions of the SED using abundances from the
ExoMol database \citep{Buldyreva2022,Guest2024}. The work from \citet{Gharib2021} predicts LiCl to be the dominant lithium-bearing molecule in an 800K atmosphere, however other molecular species (including CO$_{2}$,CH$_{4}$,NH$_{3}$, and CO) dominate the spectra for the object in this paper.  Despite the signal to noise and resolution of the data, we found no signature of LiCl.

We do, however, tentatively identify a CO$_{2}$ feature at 14--16~\micron~ present in the Spitzer IRS spectrum displayed in Figure \ref{fig:co_feature}.  All cross sections used for analysis in this work are from the (HITRAN) high-resolution transmission molecular absorption database (\citealt{Richard2012,Li2015,Hargre2020}) and the ExoMol database (\citealt{Tennyson2012,Sousa2015,Polyansky2018,Coles2019}). The JWST photometric measurements at 10, 12.8, and 18 $\micron$ as seen in Figure~\ref{fig:0415_full_sed} further corroborate the features seen in the spectrum from Spitzer.
The cross sections in Figure \ref{fig:co_feature} indicate the presence of water and ammonia in this same wavelength regime, but the narrow feature at 14--16~\micron~ is well aligned with carbon dioxide absorption. While, this $\upsilon_{2}$ band CO$_{2}$ feature has been observed in stars (e.g \citealt{Ryde1998,Justt1998}), it's the first time it has been identified in a brown dwarf atmosphere. Obtaining higher resolution spectroscopic observations of cold brown dwarfs in the mid-infrared, will allow for better constraints on the existence of this CO$_{2}$ feature and whether it is further evidence of disequilibrium chemistry. Predictions of the features observed in our SED using modern atmospheric models will be evaluated in a separate study (Suárez et al. 2024, in prep.).

\subsection{Flux Calibration Verification}

\begin{figure*}[t]
\plotone{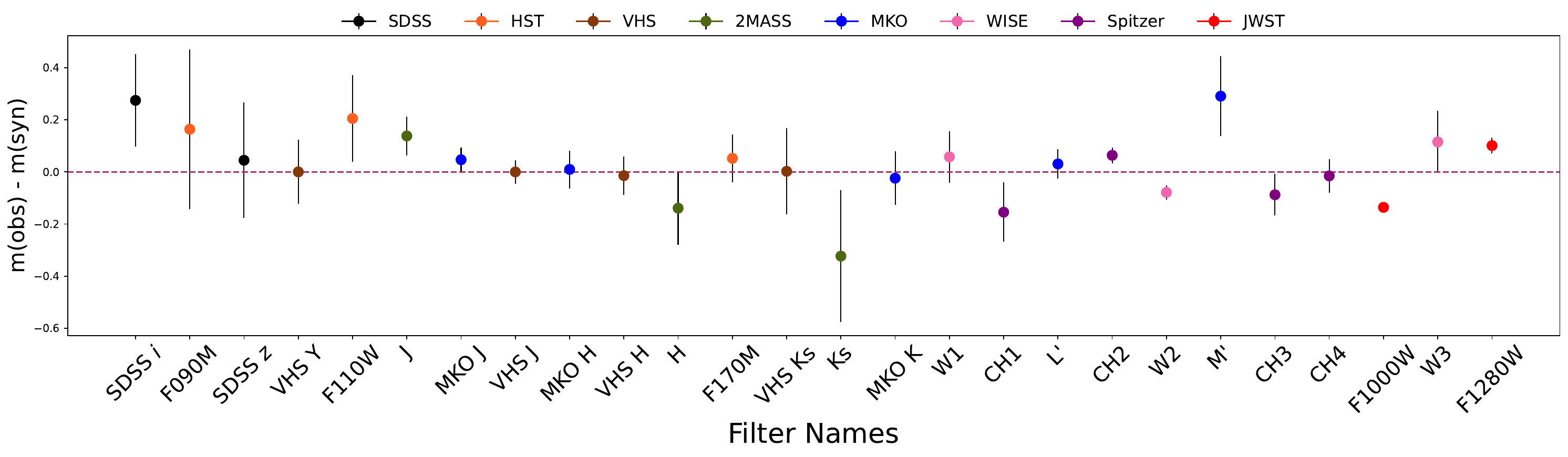}
    \caption{Residuals between observed and synthetic magnitudes from the SED. The different photometric surveys are indicated at the top and the specific filter names are at the bottom. A one to one line is indicated by the pink dashed line.\label{fig:synth_photo}}
\end{figure*}

With the flux calibrated SED for 2MASS~J0415$-$0935, we calculated synthetic photometry to evaluate the absolute flux calibration by the JWST pipeline. 
In Figure~\ref{fig:synth_photo}, we display the magnitude differences between synthetic and observed photometry from the SED. Overall the photometric points fit within 1-2 $\sigma$ of the synthetic values with a few exceptions.  

The residuals for Spitzer ch1 photometry compared to JWST synthetic points have been noted on other objects in the literature. For instance, in \citet{Luhm24} the synthetic NIRSpec spectra of WISE~J085510.83-071442.5 was compared to the Spitzer ch1 value and found to be significantly discrepant. They also highlighted a comparable divergence in the synthetic Spitzer ch1 photometry from the NIRSpec prism spectrum of WISE~J035934.06-540154.6 (\citealt{Beil2023}) and its published Spitzer value. Although, WISE~J085510.83-071442.5 and WISE~J035934.06-540154.6 are cooler Y dwarfs with increased flux in the Spitzer ch2 band, this discrepancy clearly persists in the warmer late-type T dwarfs. Furthermore, \citet{Beiler2024} corroborated this magnitude difference between synthetic and observed photometry in ch1. After analyzing 23 ultracool dwarfs, they found sources observed by JWST were systematically $\sim$0.3 mag fainter than the published Spitzer ch1 band when magnitudes were calculated synthetically using NIRSpec prism data. 

Also noteable in Figure~\ref{fig:synth_photo} is the comparison between our F1000W magnitude and the synthetic Spitzer IRS value.  The residuals between these two points are $\sim0.14$ mag (8$\sigma$) deviant.  This residual is larger than the difference of $\sim0.04$ mag reported by \citet{Beiler2024}. They tested flux calibration using MIRI LRS spectroscopic observations from 5--14~\micron in the mid-infrared region, whereas we used a Spitzer IRS spectrum from 5.2--20~\micron. The larger magnitude deviation in F1000W may be explained by differences between the Spitzer and JWST flux calibration. With upcoming JWST observations of ultra-cool dwarfs that have been previously observed with Spitzer IRS, this comparison can be further analyzed.  


\section{Conclusions}\label{CONCLU}
We present new near- and mid-infrared spectrophotometry 
for the late T dwarf 2MASS~J0415$-$0935 using JWST and the Magellan telescope. We incorporated literature data to assemble an SED spanning 0.7--20.4 $\micron$, 
which presents one of the most complete SEDs for a substellar atmosphere, covering 93\% of the flux required to calculate L$_{bol}$. 
Using the SED, we calculated a bolometric luminosity of $-5.71_{-0.01}^{+0.01}~L_\odot$ and, assuming an age range of 0.5--8.5 Gyr, we derived the following parameters: effective temperature ($729_{-10}^{+47}$ K), mass ($37_{-12}^{+10}$ $M_{\mathrm{Jup}}$), radius ($0.855_{-0.020}^{+0.110}$ $R_{\mathrm{Jup}}$ ), and $\log g$ ($5.13_{-0.35}^{+0.12}$ dex). With the JWST NIRSpec G395H spectrum we confirmed the presence of H$_{2}$O, NH$_{3}$, CH$_{4}$, CO$_{2}$ and CO.  Additionally, we demonstrated the high resolution nature of the G395H spectrum allowed for a precise radial velocity measurement of $47.1\pm1.8$ km s$^{-1}$, consistent with literature values. With the Spitzer IRS spectrum, we tentatively detected the 14 - 16 $\micron$ CO$_{2}$ feature for the first time in a substellar mass object.  

With JWST, we have been able to contextualize past characterizations of 2MASS~J0415$-$0935, and learn more about disequilibrium chemistry in some of the coldest brown dwarfs. The extensive SED presented in this work represents a unique dataset to robustly test current atmospheric models to 
learn about the chemical and physical processes that occur in these atmospheres. 

\begin{acknowledgments}
S.~A.~M. was partially funded by the John P.~McNulty Foundation for a portion of this project. J.~F. acknowledges funding support from JWST-GO-02124.001-A as well as NASA XRP Award 80NSSC22K0142 and NSF Award 1909776.
This research has made use of the Spanish Virtual Observatory (https://svo.cab.inta-csic.es) project funded by MCIN/AEI/10.13039/501100011033/ through grant PID2020-112949GB-I00.
J.~M.~V. acknowledges support from a Royal Society Science Foundation Ireland University Research Fellowship (URF$\backslash$1$\backslash$221932). This research has made use of the SVO Filter Profile Service "Carlos Rodrigo", funded by MCIN/AEI/10.13039/501100011033/ through grant PID2020-112949GB-I00.
\end{acknowledgments}

\software{SIMPLE Archive \citep{cruz_simple-astrodbsimple-db_2025}, SEDkit \citep{filippazzo_v2.0.5.sedkit_2024}, 
Astropy \citep{astropy:2022}, SVO Filters \citep{SVO2020}}
\clearpage
 

\bibliography{main}{}
\bibliographystyle{aasjournal}



\end{document}